\documentclass[aps,pre,twocolumn,showpacs]{revtex4}
\usepackage{epsfig}

\newcommand{\be}{\begin{equation}}
\newcommand{\ee}{\end{equation}}
\newcommand{\bc}{\begin{center}}
\newcommand{\ec}{\end{center}}
\newcommand{\bi}{\begin{itemize}}
\newcommand{\ei}{\end{itemize}}
\newcommand{\ba}{\begin{eqnarray}}
\newcommand{\ea}{\end{eqnarray}}
\newcommand{\ie}{{\it i.e.}}

\newcommand{\ignore}[1]{}

\begin{document}

\title{Time scale competition leading to fragmentation and recombination 
transitions\\
in the co-evolution of network and states.}

\author{Federico Vazquez}
\email[E-mail: ]{federico@ifisc.uib.es}
\author{Juan Carlos Gonz\'alez-Avella}
\author{V\'{\i}ctor M. Egu\'{\i}luz}
\author{Maxi San Miguel}

\textsc{}\affiliation{IFISC, Instituto de F\'{i}sica Interdisicplinar y 
Sistemas Complejos (CSIC-UIB), E-07122 Palma de Mallorca, Spain}
\homepage{http://ifisc.uib.es}

\date{\today}

\begin{abstract}
We study the co-evolution of network structure and node states in
a model of multiple state interacting agents. The system displays
two transitions, network recombination and fragmentation, governed 
by time scales that emerge from the dynamics.  The
recombination transition separates a frozen configuration,
composed by disconnected network components whose agents share the
same state, from an active configuration, with a fraction of links
that are continuously being rewired. The nature of this transition
is explained analytically as the maximum of a characteristic
time. The fragmentation transition, that appears between two
absorbing frozen phases, is an anomalous order-disorder
transition, governed by a crossover between the time scales that
control the structure and state dynamics.
\end{abstract}

\pacs{89.75.Fb; 05.65.+b; 89.75.-k; 64.60.Cn}

\maketitle

\section{Introduction} Recent findings in the topological
characterization of real complex networks have triggered a
theoretical understanding of diverse complex systems. A great deal
of effort has been devoted to the modelling of complex networks
from a topological point of view, and to the dynamics on different
classes of networks \cite{Albert02}. In the latter, the evolution
of the states of the nodes is usually assumed much faster that the
time characterizing the network dynamics. Less effort has been
devoted to the understanding of the entangled co-evolution of
network structure and state dynamics, \ie, how the structure of a
network affects the dynamics on it and vice versa. For instance in
a social system, individuals shape their opinions depending on
their neighbors' opinions, but simultaneously, the individuals'
opinions affect with whom they interact
\cite{Skyrms00,Zimmermann04,Marsili04,Eguiluz05b,Gil06,Holme07}.
The network of interactions co-evolve with the dynamics of
opinions.

The influence of network topology has been analyzed extensively
for several models of consensus formation
\cite{Suchecki05,Sanmiguel05}. A main question is related to the
mechanisms and network topologies that lead to consensus, \ie, all
agents having the same state. Based on social pressure, how state
changes, and homophily, the tendency of individuals to interact with similar 
others, the Axelrod model 
\cite{Axelrod97} represents a paradigmatic model
displaying an order-disorder nonequilibrium transition
\cite{Castellano00}. When the network of interaction is regular,
random or small-world, the system orders if the degree of initial
disorder is below a critical value
\cite{Castellano00,Klemm03a,Klemm03b,Vazquez07}. In the ordered
phase, a domain (set of connected agents with the same state) of
the order of the system size spans the network, while in the
disordered phase, many small domains are formed.
Motivated by the mechanisms of social pressure and homophily, in
this paper we analyze the co-evolution in the Axelrod model
between the network of interactions and the state dynamics of the
nodes, by allowing interaction links to be rewired depending on
the state of the agents at their ends.

\section{The model} A population of $N$ agents are located at the
nodes of a network. The state of an agent $i$ is represented by an
$F$-component vector $\sigma_{if}$, $f=1,2,\ldots,F$, and $i=1,2,
\ldots, N$, where each component represents an agent's attribute.
There are $q$ different choices or traits per feature, labelled
with an integer $\sigma_{if} \in \{0,\ldots,q-1\}$, giving raise
to $q^F$ possible different states.

Initially agents take one of the $q^F$ states at random. In a time step an
agent $i$ and one of its neighbors $j$ are randomly chosen:

\begin{enumerate}
\item If the agents share $m>0$ features, they interact
    with probability equal to the fraction of shared features,
  i.e., the overlap ($m/F$). In case of interaction, an unshared
  feature is selected at random and $i$ copies $j$'s value for
  this feature.

\item If the agents do not share any feature, then $i$
    disconnects its link to $j$ and connects it to a randomly
  chosen agent that $i$ is not already connected to.
\end{enumerate}

Step $1$ describes the original Axelrod dynamics: Alike agents
become even more similar as they interact, increasing the
probability of future interaction. Step $2$ implements the network
co-evolution: {\em incompatible agents}, i.e., agents with no
features in common, tend to get disconnected. We have performed
extensive numerical simulations to study the behavior of the
system as the control parameter $q$ is varied, for different
population sizes $N$, number of features $F=3$, and starting from
a random network with average degree $\langle k \rangle = 4$. The
results do not depend on the initial network topology, because the
repeated rewiring dynamics leads to a random network with a
Poisson degree distribution.

The model displays two transitions, both very different in nature.
The first one is an order-disorder transition at $q=q_c$ between
two frozen phases, associated with the fragmentation of the
network. The dynamics leads to the formation of network components, 
where a component is a set of connected nodes.
In a frozen configuration agents that belong to the same component have 
the same state.
For $q<q_c$
(ordered phase I), the system reaches a configuration composed by
a giant component of the order of the system size, and a set of
small components; for $q>q_c$ (disordered phase II) the large
component disintegrates in many small disconnected components. The
second transition, related with network recombination, occurs at
$q=q^*$ between the phase II and an active phase III, where the
system reaches a dynamic configuration with links that are
permanently rewired.

\begin{figure}[t]
\begin{center}
 \vspace*{0.cm}
 \includegraphics*[width=0.45\textwidth]{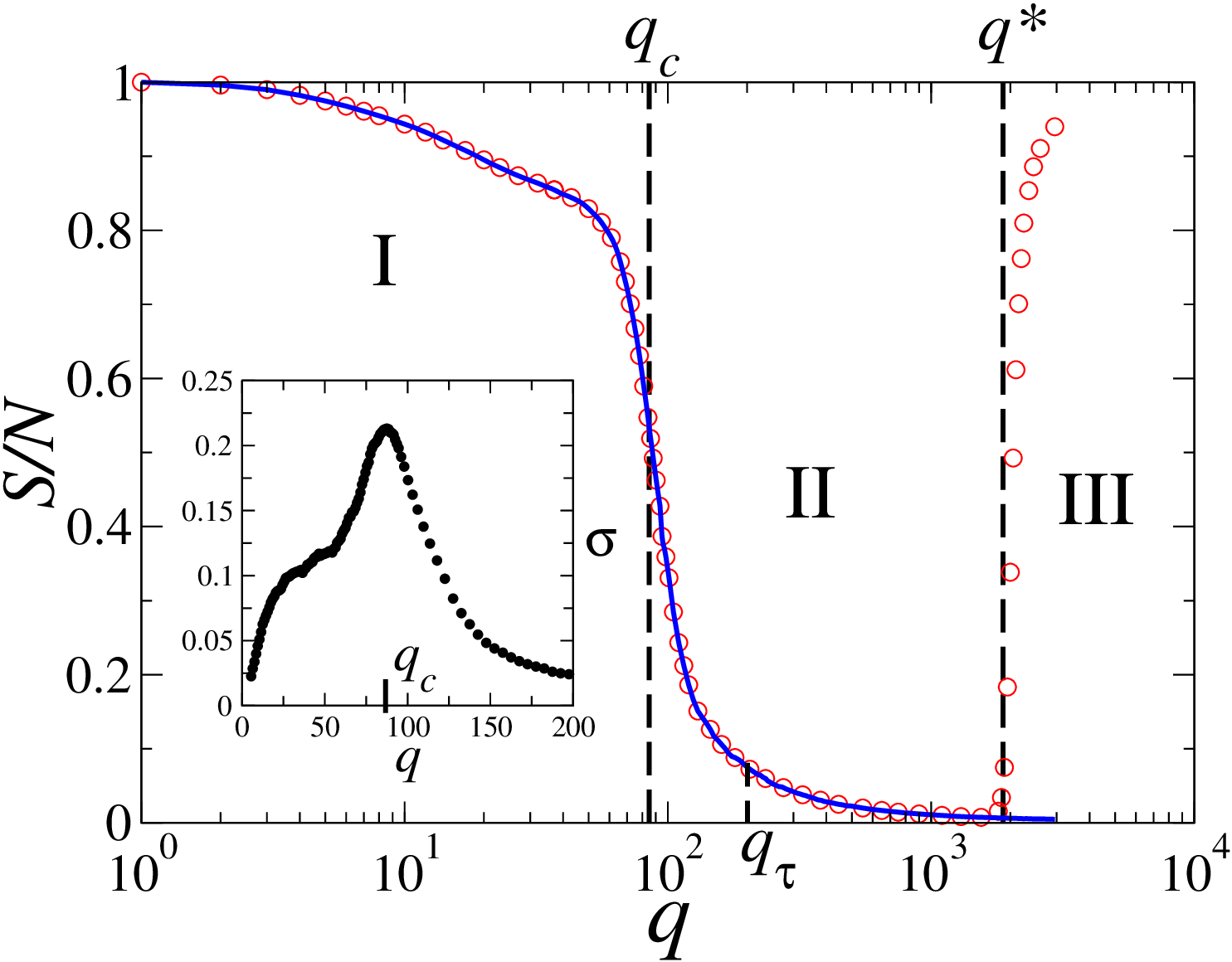}
 \caption{Average relative size of the largest network component (circles) and
   largest domain (solid line) in the stationary configuration vs $q$, for
   $N=2500$, averaged over $400$ realizations. The vertical lines at
   $q_c=85$ and $q^*=1875$ indicate the transition points between the
   different phases. Inset: fluctuations are maximum at the critical point
   $q_c$.}
 \label{Sq}
\end{center}
\end{figure}

\section{Frozen phases (I and II)} For small values of $q$ in phase
I, the average size of the largest network component $S$ in the
final configuration is of the order of the system size $N$
(Fig.~\ref{Sq}), due to the high initial overlap between the
states of neighboring agents. As $q$ increases inside phase I, the
initial overlap decreases and $S$ also slowly decreases. For
larger values of $q$ (phase II), many distinct domains are formed
initially inside the components which break into many small
disconnected components and, as a result, $S$ reaches a value much
smaller than $N$ (network fragmentation). During the evolution of
the system, a network component can have more than one domain.
However, in the final configuration of phases I and II, one
component corresponds to one domain.

\begin{figure}[t]
\begin{center}
 \vspace*{0.cm}
 \includegraphics*[width=0.45\textwidth]{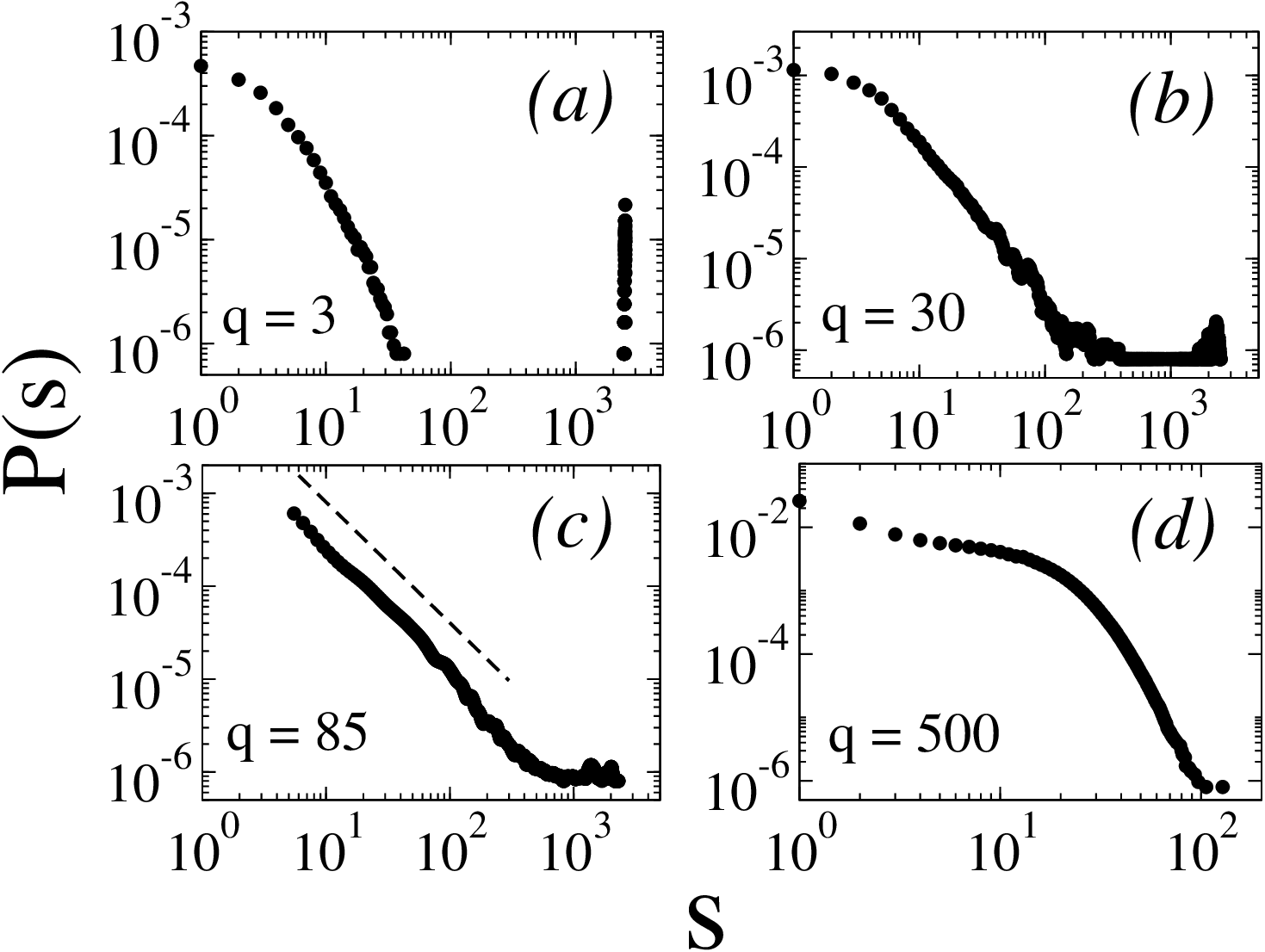}
 \caption{Size distribution of network components for $N=2500$
 and values of $q$ (a,b) below,
 (c) at, and (d) above the transition point $q_c \simeq 85$.
 The dashed line represents a power law with exponent $-1.3 \pm 0.02$.}
 \label{P-s}
\end{center}
\end{figure}

The transition point from phase I to phase II is defined by the value $q=q_c$
for which the fluctuations in $S$ reach a maximum value. This value
corresponds to the point where the order parameter $S$ suffers a sudden
drop. For $N=2500$ we find $q_c = 85 \pm 2$ (see Fig.~\ref{Sq}). To 
further
investigate this transition point we calculated the size distribution of
network components $P(s)$ (see Fig.~\ref{P-s}). For small $q$, $P(s)$ shows a
peak that corresponds to the average size of the largest component $S$, and
has an exponential decay corresponding to the distribution of small
disconnected components. The peak at $S$ decreases as $q$ increases, giving
raise to a power law decay of $P(s)$ at $q_c \simeq 85$, a signature of a
transition point \cite{Stauffer92,Klemm03a}.

\begin{figure}[t]
\begin{center}
 \vspace*{0.cm}
 \includegraphics*[width=0.45\textwidth]{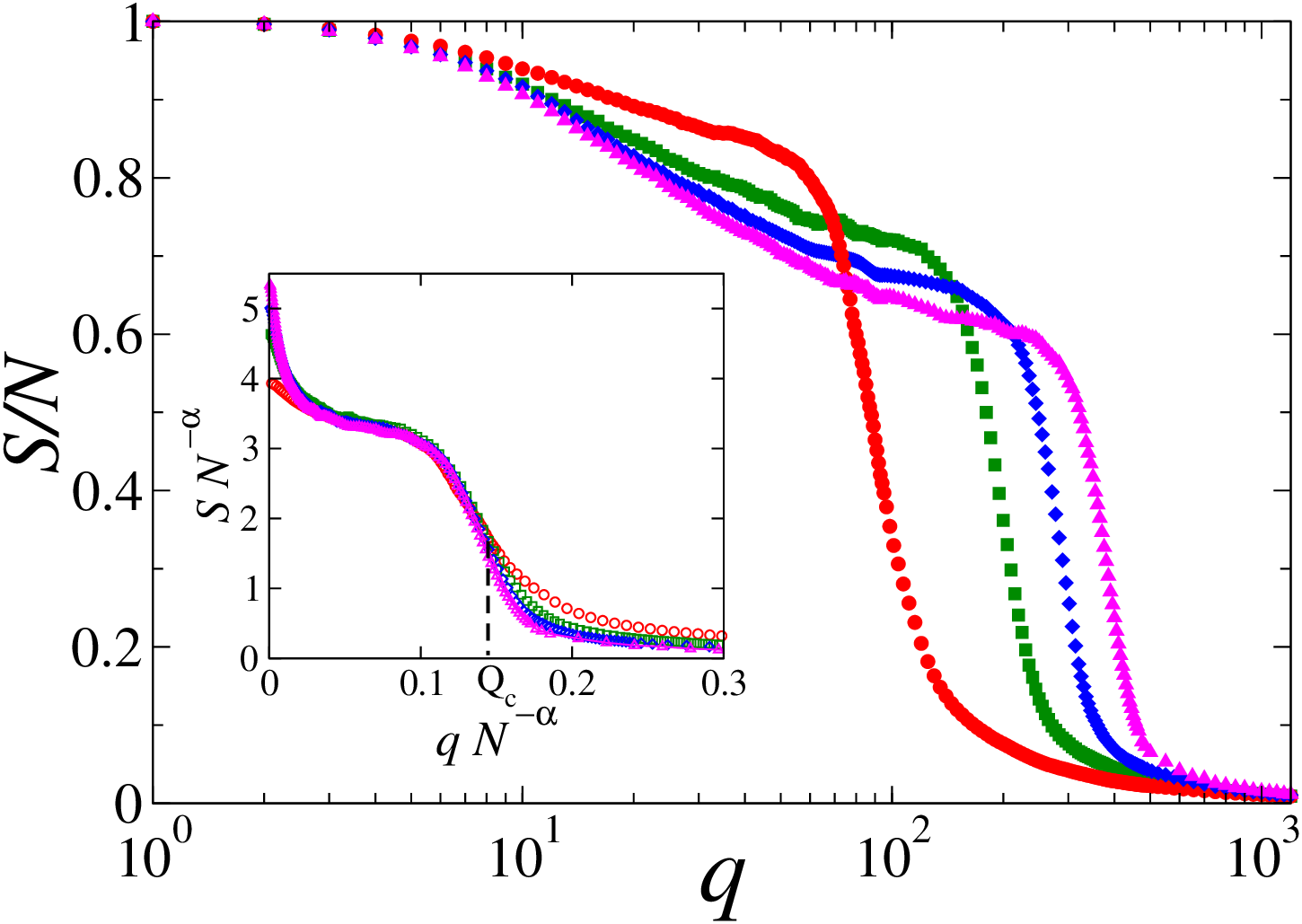}
 \caption{Average relative size of the largest network component
 $S/N$ vs $q$ for system sizes $N= 2500$, $6400$, $10000$ and $14400$
 (left to right), averaged over $400$ realizations. Inset: Finite size scaling.
 The data collapses below the scaled transition point $Q_c = q_c
 N^{-\alpha}$, with $\alpha=0.82$.}
\label{Sq1}
\end{center}
\end{figure}

The behavior of the order parameter $S/N$ depends on system size
(Fig.~\ref{Sq1}). When both axis are rescaled by $N^{-\alpha}$
with $\alpha = 0.82 \pm 0.01$ the data collapses for $q$ smaller than
$q_c$. This implies that $q_c$ increases with the system size $N$
as $q_c \sim N^{\alpha}$, and suggests a scaling relation for $S$
in the ordered phase $S = N^{\alpha}f(N^{-\alpha} q)$, where
$f(\cdot)$ is a scaling function. The scaling relation implies
that the discontinuity disappears in the large $N$ limit.
Thus, not only the transition point diverges as $q_c \sim
N^{~0.82}$ but also the amplitude of the order parameter $S/N \sim
N^{-0.18}$ vanishes as $N$ goes to infinity.

These results, maximum of the fluctuations, data collapse and distribution of
component sizes, identify $q_c$ as the critical point of the transition. The
fact that the exponent of the size distribution is smaller than $2$ and that
the discontinuity of the order parameter tends to $0$ as system size
increases, suggest that in the large $N$ limit the transition becomes
continuous with $q_c \to \infty$.

\section{Active phase (III)} We analyze this phase by looking at the rewiring 
dynamics. A link that
connects a pair of incompatible agents is randomly rewired until
it connects two compatible agents, i.e., agents with at least one
feature in common. If the number of pairs of compatible agents
$L_c$ is larger than the total number of links in the system
$\langle k \rangle N/2$, this rewiring process continues until all
links connect compatible agents. Later, the system evolves until
each component constitutes a single domain, the frozen configurations
reached in phases I and II. If, on the contrary, $L_c$ gets
smaller than $\langle k \rangle N/2$, the system evolves
connecting first all $L_c$ pairs by links. The state dynamics
stops when no further change of state is possible, but the
rewiring dynamics continues for ever with approximately $\langle k
\rangle N/2 - L_c$ links that repeatedly fail to attach compatible
agents. This is the active configuration observed in phase III.
Thus, in contrast to phases I and II, in the stationary
configurations of phase III there are typically more than one
domain per component. The size of the largest component abruptly
increases at $q^*$ indicating that a giant component reappears
(network recombination), while the size of the largest domain
continues decreasing (see Fig.~\ref{Sq}).

To estimate $q^*$, we will assume $L_c$ constant during the evolution as for
$q$ large the state of the agents does not evolve much. Thus, with the ansatz
$L_c \simeq L_c(t=0) \simeq \frac{N(N-1)}{2} \left[ 1-(1-1/q)^F\right] \simeq
N^2 F/2 q$, for $q \gg F$ and \\ $N \gg 1$, the condition at the transition
point is $N^2 F/2 q^* \simeq \langle k \rangle N/2$, so that
\begin{equation}
q^* \simeq \frac{N F}{\langle k \rangle}~
\end{equation}
is the value of the recombination transition point. For the values considered
here we obtain $q^* = 1875$ in good agreement with the numerical results
(Fig.~\ref{Sq}).

\section{Dynamic time scales} 
\begin{figure}[t]
\begin{center}
 \vspace*{0.cm}
 \includegraphics*[width=0.45\textwidth]{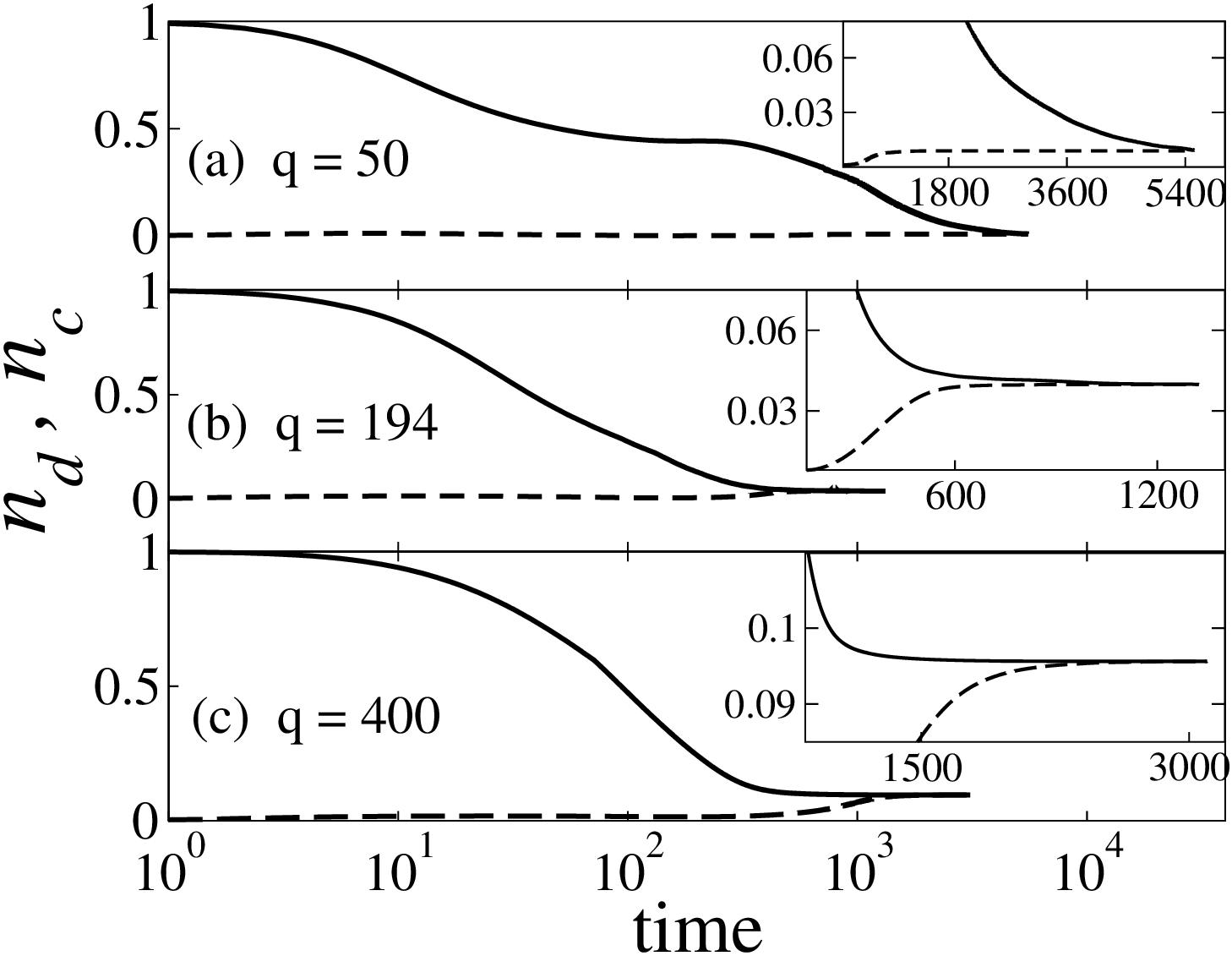}
 \caption{Time evolution of the density of domains $n_d$ (solid line)
 and network components $n_c$ (dashed line), for $N=2500$ and
 values of $q$ (a) below, (b) at, and (c) above the transition
 point $q_\tau = 194 \pm 10 $. Each inset is a zoom of the region that
 shows the approach to the final configuration.}
\label{Ng-2500-t}
\end{center}
\end{figure}

\begin{figure}[t]
\begin{center}
 \vspace*{0.cm}
 \includegraphics*[width=0.45\textwidth]{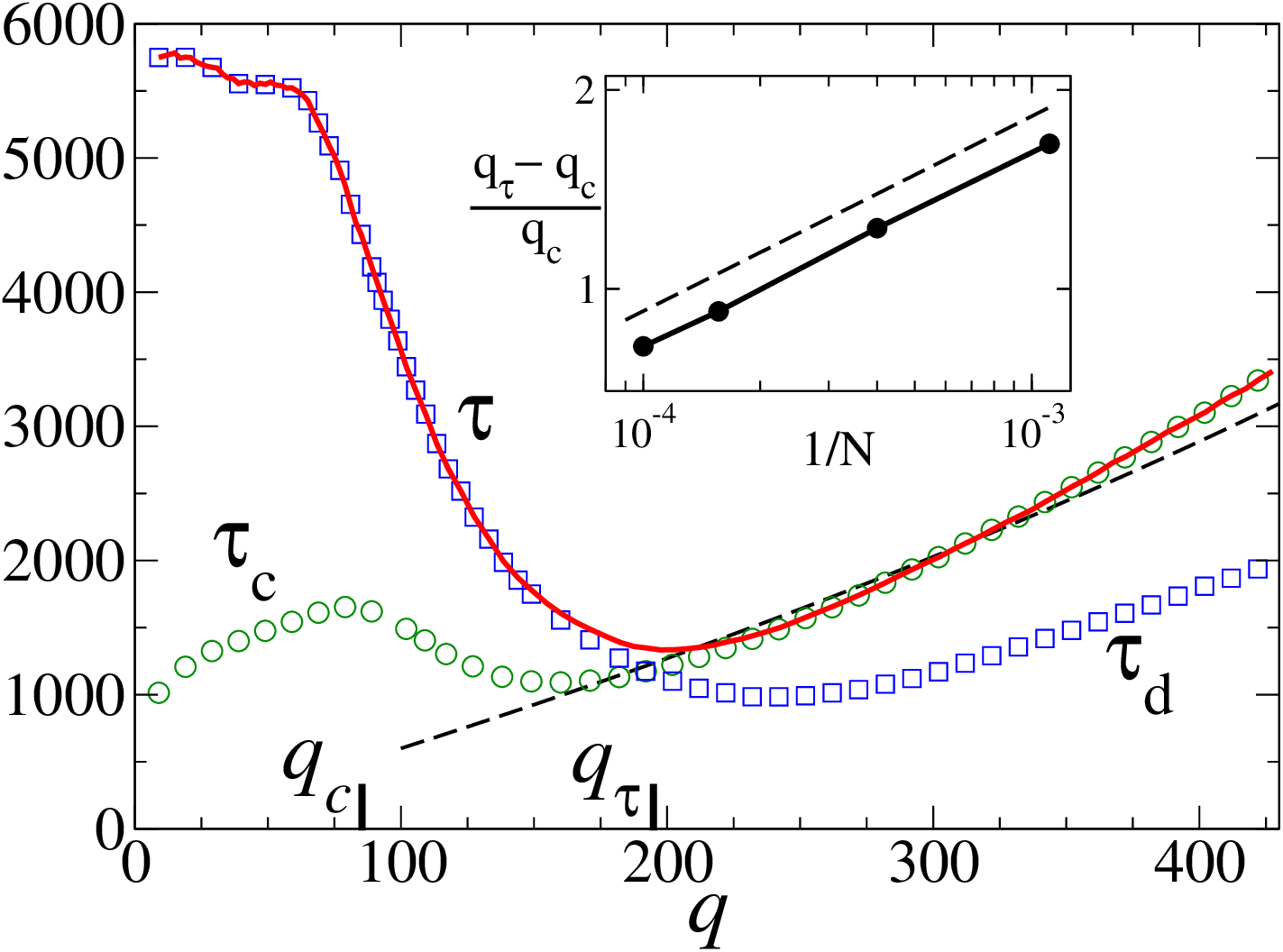}
 \caption{Convergence times $\tau_c$ (circles), $\tau_d$ (squares) and $\tau$
 (solid line) vs $q$ for $N=2500$ and averaged over $500$ configurations. The
 result from Eq.~(\ref{tau}) (dashed line) is compared with $\tau$ and $\tau_c$
 for $q>q_\tau = 194 \pm 10$. Inset: relative difference
 between the transition points $q_\tau$ and $q_c$ vs $1/N$ in log-log scale.
 The dashed line has slope $0.30 \pm 0.01$.}
\label{T-N2500}
\end{center}
\end{figure}

In order to understand the final
structure of the network in phases I and II we analyzed the time
evolution of the nodes' states and interaction links. In
Fig.~\ref{Ng-2500-t} we plot the time evolution of the density of
network components, $n_c=$ number of components$/N$, and domains,
$n_d=$number of domains$/N$, averaged over $1000$ realizations,
and for three values of $q$. An interesting quantity is the
average time to reach the final frozen configuration $\tau$. If
$\tau_d$ [$\tau_c$] is the average time at which $n_d$ [$n_c$]
reaches its stationary value, then $\tau$ is largest between
$\tau_d$ and $\tau_c$ (see Figs.~\ref{Ng-2500-t} and
\ref{T-N2500}). The curves for $\tau_d$ and $\tau_c$ as a function
of $q$ cross at a value $q_\tau$ (Fig.~\ref{T-N2500}). As we shall
see, $q_\tau$ identifies a transition between two different dynamic
regimes for the formation of domains and network components, that lead 
to the frozen configurations observed in phases I and II 
(Figs.~\ref{Ng-2500-t} and \ref{T-N2500}).

For $q<q_\tau$, the dynamics causes the network to break into a giant 
component and small components. Due to the initial overlap between the states 
of the
agents inside each component, the network stops evolving at a time
$\tau_c$ where $n_c$ reaches its stationary maximum value (see
Fig.~\ref{Ng-2500-t}a). After this stage, domains compete inside
each component, until only one domain occupies each component. The
approach to the frozen configuration is controlled by the
coarsening process inside the largest component, whose structure
is similar to a random network due to the random rewiring
dynamics. Then, given that the dynamics of the last and longest
stage before reaching consensus inside this component is governed by 
interfacial noise as in the voter model \cite{Klemm03c}, $\tau$ is expected 
to scale as the size of the largest component $\tau \sim S$ \cite{Sood05}.

For $q>q_\tau$, there is a transient during which $n_d$ decreases,
indicating that, in average, domains grow in size (see
Fig.~\ref{Ng-2500-t}c). At time $\tau_d$, $n_d$ reaches a
stationary value when the overlap between distinct domains
is zero. At this stage domains are still interconnected by links
between incompatible agents. As domains progressively disconnect
from each other $n_c$ increases. When finally the links connecting
incompatible agents disappear all domains get fragmented, $n_c$
equals $n_d$ and the system reaches its final configuration in a
time $\tau = \tau_c$.

At $q_\tau$, the time scale governing the state dynamics is the same as the
time scale governing the network dynamics $\tau_d = \tau_c$
(Figs.~\ref{Ng-2500-t}b,\ref{T-N2500}). Thus, it indicates that the ordered
phase I is dominated by a slower state dynamics on a network that freezes on a
fast time scale, while in the disordered phase II the state dynamics freezes
before the network reaches a final frozen configuration. Even though
$q_{\tau}$ is found to be larger than the critical point $q_c$, the relative
difference between $q_\tau$ and $q_c$ decreases with $N$ (inset of
Fig.~\ref{T-N2500}), suggesting that both transition points become equivalent
in the large $N$ limit. Thus, the competition between the time scales
$\tau_d$ and $\tau_c$ governs the fragmentation transition at $q_c$.

In the remainder, we derive an approximate expression for $\tau$ in phase II
by studying the decay of the number of links between incompatible agents
$N_0$. We shall see that this approach unveils the transition from the frozen
to the active phase, leading to the transition point $q^*$.

In the continuum time limit, and neglecting the creation of
incompatible links, $N_0$ decays according to the equation: \be
\frac{d N_0}{d t} \simeq - \frac{1}{\frac{1}{N}}
\frac{2N_0}{\langle k \rangle N} \frac{N_c}{N} = - \frac{2 N_0 \;
N_c}{\langle k \rangle N}. \label{dN0dt} \ee In a time step
$\Delta t=\frac{1}{N}$, an incompatible link $(i,j)$ is chosen
with probability $\frac{2N_0}{\langle k \rangle N}$. One of its
ends $j$ is moved to a random node $k$. The probability that $k$
is compatible to $i$ is $\frac{N_c}{N}$, where $N_c$ is the number
of compatible agents to $i$ but still not connected to $i$. In a
mean-field spirit, every node has $\langle k \rangle$ edges that
need to be connected to $\langle k \rangle$ different compatible
agents. We approximate $N_c$ as the average number of compatible
agents per agent. When $i$ attaches an edge to a compatible agent,
$N_0$ is reduced by one while $N_c$ is reduced by $2/N$ given that
both $i$ and $k$ loose a compatible partner. Assuming that the set
of compatible agents to $i$ remains the same we write $N_c \simeq
\frac{2 N_0}{N} + A$, where $A = N_c(0) - \frac{2 N_0(0)}{N}
\simeq \frac{N F}{q}-\langle k \rangle$ for $N \gg 1$ and $q \gg
F$. Substituting this last expression for $N_c$ into
Eq.~(\ref{dN0dt}) and rewriting it in terms of the transition
point $q^* = NF/\langle k \rangle$ we obtain \be \frac{d N_0}{d t}
\simeq - \frac{2 N_0}{\langle k \rangle N} \left( \frac{2 N_0}{N}
+ \frac{\langle k \rangle}{q} \left( q^*-q \right) \right).
\label{dN0dt1} \ee Equation~(\ref{dN0dt1}) has two stationary
solutions. For $q<q^*$, the steady configuration is $N_0^S=0$,
corresponding to the frozen phases I and II, while for $q>q^*$ the
stationary solution $N_0^S=\frac{\langle k \rangle N}{2q}(q-q^*)$
corresponds to the active phase III.  We recover our previous result 
that for $q>q^*$ the system reaches a stationary configuration 
with a constant fraction of links $N_0$ larger than zero, that are 
permanently rewired.  Therefore, the system never freezes.
Note that in the limit of very large $q$, all agents are initially 
incompatible, consequently
$N_0$ approaches to the total number of links $\frac{1}{2} \langle k
\rangle N$. 

Integrating Eq.~(\ref{dN0dt1}) by a partial fraction
expansion gives \be t = \frac{q N}{2(q^*-q)} \ln \left[
\frac{q+\frac{\langle k \rangle N}{2 N_0(t)}(q^*-q)}{q^*} \right].
\ee 
For $q<q^*$, the system freezes at a time $\tau$ at which $N_0 \simeq 1$,
thus 
\be \tau \simeq \frac{q N}{2(q^*-q)} \ln \left[
\frac{q+\frac{1}{2}\langle k \rangle N(q^*-q)}{q^*} \right]~,
\mbox{~for~} q_\tau < q < q^*. \label{tau} 
\ee 
This result is in agreement with the numerical solution 
(Fig.~\ref{T-N2500}).

For $q>q^*$, the system reaches a stationary configuration.  Thus we define 
$\tau$ as the time at which 
$N_0 \simeq \frac{\langle k \rangle N}{2q}(q-q^*) +1$, then  
\be 
\tau \simeq \frac{q N}{2(q^*-q)} \ln \left[
\frac{2 q^2}{q^*(\langle k \rangle N(q-q^*)+2q)} \right]~,
\label{taus} 
\ee 
for $q^* < q < \frac{N^2 F}{2}$. $\tau$ 
decreases with $q$ and it vanishes for  $q > N^2 F/2$ where initially 
all pairs of nodes are incompatible, thus the system starts from a stationary 
configuration, giving $\tau = 0$ for $q > \frac{N^2 F}{2}$. 

From Eqs.~(\ref{tau}) and ~(\ref{taus}) we obtain that $\tau$  reaches a
maximum value equal to $\frac{1}{4} \langle k \rangle N^2$ at $q=q^*$; an
indication of the transition.

\section{Summary and Conclusions} We have studied the Axelrod model with 
co-evolution of the interaction network
and state dynamics of the agents. The interplay between structure
and dynamics gives rise to two different transitions. First, a
recombination transition between a frozen and an
active phase.  The characteristic time to reach a stationary configuration
shows a maximum at the transition point between these two phases. 
Second, an order-disorder transition associated
with network fragmentation that appears at a critical value $q_c$
where the component size distribution follows a power-law. Finite
size scaling analysis suggests that in the large $N$ limit this
transition becomes continuous with $q_c$ going to infinity. The
fragmentation is shown to be a consequence of the competition
between two coupled mechanisms, network formation and state
formation. These mechanisms are governed by two internal time
scales, $\tau_c$ and $\tau_d$ respectively, which are not
controlled by external parameters, but they emerge from the
dynamics. For $q<q_c$ the network components, that are formed
first, control the formation of states.  For $q>q_c$ the fast
formation of domains shape the final structure of the network.

An important aspect of the co-evolution is that the network evolution
is coupled to the state of the agents, in contrast to other models
where the links are severed, reconnected and/or appear as a random
process independent of the state of the nodes.
The robustness of the fragmentation and
recombination in the presence of noise
should also be considered \cite{Centola06}. Our results provide a
simple mechanism, co-evolution, that could explain the community
structure found in general in the analysis of complex networks,
and in particular of social systems \cite{Palla07} where this
mechanism can be understood in terms of network homophily
\cite{Centola06}.

\begin{acknowledgments}
We acknowledge fruitful discussions with Damon Centola and
financial support from the MEC (Spain) through projects CONOCE2
(FIS2004-00953) and SICOFIB (FIS2006-09966).
\end{acknowledgments}



\begin{thebibliography}{}

\bibitem{Albert02} R. Albert and A.-L. Barab\'asi, Rev. Mod. Phys. {\bf 74},
47 (2002).

\bibitem{Skyrms00} B. Skyrms and R. Pemantle, Proc. Natl. Acad. Sci. USA {\bf
97}, 9340 (2000).

\bibitem{Zimmermann04} M.G. Zimmermann, V.M. Egu\'{\i}luz, and M. San Miguel,
Phys. Rev. E {\bf 69}, 065120(R) (2004).

\bibitem{Marsili04} M. Marsili, F. Vega-Redondo, and F. Slanina,
Proc. Natl. Acad. Sci. USA {\bf 101}, 1439 (2004).

\bibitem{Eguiluz05b} V.M. Egu\'{\i}luz, M.G. Zimmermann, C.J. Cela-Conde, and
M. San Miguel, Am. J. Sociol. {\bf 110}, 977 (2005).

\bibitem{Gil06} S. Gil and D.H. Zanette, Phys. Lett. A {\bf 356}, 89 (2006).

\bibitem{Holme07} P. Holme and M.E.J. Newman, Phys. Rev. E {\bf 74}, 056108
(2006).

\bibitem{Sanmiguel05} M. San Miguel, V.M. Egu\'{\i}luz, R. Toral, and
K. Klemm, Comput. Sci. Eng. {\bf 7}, 67 (2005).

\bibitem{Suchecki05} K. Suchecki, V.M. Egu\'{\i}luz, and M. San Miguel,
Europhys. Lett. 69 228 (2005);Phys. Rev. E {\bf 72}, 036132 (2005).

\bibitem{Axelrod97} R. Axelrod, J. Conflict Res. {\bf 41}, 203 (1997)

\bibitem{Castellano00} C. Castellano, M. Marsili, and A. Vespignani,
Phys. Rev. Lett. {\bf 85}, 3536 (2000).

\bibitem{Klemm03b} K. Klemm, V.M. Egu\'{\i}luz, R. Toral, and M. San Miguel,
Phys. Rev. E {\bf 67}, 026120 (2003).

\bibitem{Klemm03a} K. Klemm, V.M. Egu\'{\i}luz, R. Toral, and M. San Miguel,
Physica A {\bf 327}, 1 (2003).

\bibitem{Vazquez07} F. Vazquez and S. Redner, Europhys. Lett. {\bf 78}, 18002
(2007).

\bibitem{Stauffer92} D. Stauffer and A. Aharony, {\sl Introduction to
percolation theory} (Taylor and Francis, London, 1992).

\bibitem{Klemm03c} K. Klemm, V.M. Egu\'{\i}luz, R. Toral, and M. San Miguel,
Phys. Rev. E {\bf 67}, 045101(R) (2003).

\bibitem{Sood05} V. Sood and S. Redner, Phys. Rev. Lett. {\bf 94} 178701
(2005).

\bibitem{Centola06} D. Centola, J.C. Gonz\'alez-Avella, V.M. Egu\'{\i}luz, and
M San Miguel, {\tt physics/0609213}.

\bibitem{Palla07} G. Palla, A.-L. Barab\'asi, and T. Vicsek, Nature (London)
 {\bf 446}, 664 (2007).


\end{thebibliography}
\end{document}